\documentclass[conference]{IEEEtran}
\IEEEoverridecommandlockouts

\usepackage{pifont}
\newcommand{\cmark}{\ding{51}}%
\newcommand{\xmark}{\ding{55}}%
\usepackage{amsmath}
\usepackage{amssymb}
\usepackage{amsfonts}
\usepackage{graphicx}
\usepackage{textcomp}
\usepackage{xcolor}
\usepackage{multirow}
\usepackage{soul}
\usepackage{tikz}
\usepackage{booktabs}
\usepackage{algorithm}
\usepackage{algpseudocode}
\usepackage{colortbl}
\usepackage{bm}
\usepackage{float}
\usepackage{filecontents}
\usepackage{CJK}
\usepackage{tikz}
\usepackage{makecell}
\usepackage{algorithm}
\usepackage{algorithmicx}
\newcommand\encircle[1]{%
\tikz[baseline=(X.base)] 
  \node (X) [draw, scale=0.75, shape=circle, inner sep=0, fill=black, text=white, minimum size=0em] {\strut #1};}

\usepackage{amsmath}
\usepackage{amssymb}
\usepackage{amsfonts}
\usepackage{graphicx}
\usepackage{textcomp}
\usepackage{xcolor}
\usepackage{multirow}
\usepackage{soul}
\usepackage{tikz}
\usepackage{booktabs}
\usepackage{algorithm}
\usepackage{algpseudocode}
\usepackage{colortbl}
\usepackage{bm}
\usepackage{float}
\usepackage{filecontents}
\usepackage{CJK}
\usepackage{makecell}
\usepackage{algorithm}
\usepackage{algorithmicx}
\usepackage[most]{tcolorbox}
\def\BibTeX{{\rm B\kern-.05em{\sc i\kern-.025em b}\kern-.08em
    T\kern-.1667em\lower.7ex\hbox{E}\kern-.125emX}}
\begin{document}

\title{EIM-TRNG: Obfuscating Deep Neural Network Weights with Encoding-in-Memory True Random Number Generator via RowHammer
\vspace{-0.7em}}

\author{Ranyang Zhou$^\dagger$, Abeer Matar A. Almalky$^\ddagger$, Gamana Aragonda$^\dagger$, Sabbir Ahmed$^\ddagger$, Filip Roth Trønnes-Christensen$^\dagger$,\\ Adnan Siraj Rakin$^\ddagger$, and Shaahin Angizi$^\dagger$  \\
\small $^\dagger$Department of Electrical and Computer Engineering, New Jersey Institute of Technology, Newark, NJ, USA\\
$^\ddagger$Department of Computer Science, State University of New York at Binghamton, NY, USA\\
rz26@njit.edu, arakin@binghamton.edu, shaahin.angizi@njit.edu \vspace{-1em}
\\}
\maketitle

\begin{abstract}
True Random Number Generators (TRNGs) play a fundamental role in hardware security, cryptographic systems, and data protection. In the context of Deep Neural Networks (DNNs), safeguarding model parameters—particularly weights—is critical to ensure the integrity, privacy, and intellectual property of AI systems. While software-based pseudo-random number generators are widely used, they lack the unpredictability and resilience offered by hardware-based TRNGs. In this work, we propose a novel and robust Encoding-in-Memory TRNG called EIM-TRNG that leverages the inherent physical randomness in DRAM cell behavior, particularly under RowHammer-induced disturbances, for the first time. We demonstrate how the unpredictable bit-flips generated through carefully controlled RowHammer operations can be harnessed as a reliable entropy source. Furthermore, we apply this TRNG framework to secure DNN weight data by encoding via a combination of fixed and unpredictable bit-flips. The encrypted data is later decrypted using a key derived from the probabilistic flip behavior, ensuring both data confidentiality and model authenticity. Our results validate the effectiveness of DRAM-based entropy extraction for robust, low-cost hardware security and offer a promising direction for protecting machine learning models at the hardware level.
\end{abstract}

\section{Introduction}
Recently, DNNs have emerged as a popular solution to perform a wide range of real-world tasks, including but not limited to language processing and vision ~\cite{koppula2019eden}. Training these large-scale DNN models requires an enormous amount of training resources and financial support, making them an important intellectual property (IP) for the model developer. In addition, protecting model parameters improves their security, as prior works have shown that having access to model architecture and weights allows an adversary to launch strong white-box attacks on leaked DNN model~\cite{goodfellow2014explaining,rakin2022deepsteal,zhou2025compromising}. Hence, protecting internal model information, such as topology and weights, is critical to protect both financial and security losses for model developers.

TRNGs play a critical role in protecting the privacy of DNNs~\cite{9116243,liu2024tbnet,liu2023mirrornet}, especially in scenarios where model parameters need to be stored or transmitted securely. Randomness is fundamental to cryptographic protocols and privacy-preserving techniques, such as secure multiparty computation and watermarking \cite{mohassel2017secureml,uchida2017embedding}, which rely on non-deterministic keys or initialization vectors to prevent reverse engineering, model piracy, or unauthorized inference. While Pseudo-Random Number Generators (PRNGs) can generate sequences that appear random, they are ultimately deterministic and vulnerable to attacks if the initial seed or algorithm is known. In contrast, TRNGs derive randomness from physical entropy sources, ensuring unpredictability and robustness against such threats.

However, generating high-quality TRNs in hardware is a non-trivial task \cite{wallace2001survey}. Conventional TRNGs often rely on specialized circuits to exploit sources such as thermal noise, metastability, or jitter \cite{sunar2007cmos,dichtl2007true}. These approaches, while effective, typically demand custom circuitry, increasing design complexity and hardware cost, and are not always feasible for deployment in commodity devices. To address these limitations, researchers have explored DRAM as an alternative entropy source due to its ubiquity, low cost, and physical behavior that can be exploited without hardware modification. Several notable efforts in this direction include DRNG, DTRNG, FracDRAM, and QUAC-TRNG~\cite{eckert2017drng,humood2021dtrng,gao2022fracdram,olgun2021quac}.
DRNG~\cite{eckert2017drng} utilizes unpredictable startup values of DRAM cells. Upon powering up, DRAM cells show random values due to floating bitlines and manufacturing variations. DTRNG~\cite{humood2021dtrng} introduces randomness by performing weak write operations, applying reduced wordline (WL) voltage to generate partial charge transfers. This results in varied output due to process noise, and it requires no hardware changes. FracDRAM~\cite{gao2022fracdram} introduces the Frac operation, a sequence of Activate (\texttt{ACT}) and Precharge (\texttt{PRE}) commands to leave the cells in a metastable voltage state near $\frac{V_{DD}}{2}$. This uncertainty is then sampled by the sense amplifiers, yielding random results. Similarly, QUAC-TRNG~\cite{olgun2021quac} leverages a QUAC operation to activate and interrupt multiple rows simultaneously, inducing charge sharing and generating unpredictable voltages that produce random bit values after sensing.

Despite these advancements, most DRAM-based TRNGs are designed for continuous randomness extraction or entropy pooling \cite{dichtl2007true,sunar2007cmos,zhou2023threshold}. In this work, we explore a different direction by leveraging DRAM behavior to generate a one-time-use key suitable for encrypting a memory page. This is particularly useful for secure watermarking or encrypting the weights of a DNN model in a way that ensures confidentiality and traceability, while still enabling recovery by authorized users. To achieve this, we propose \textbf{\textit{EIM-TRNG}}, a novel Encoding-in-Memory TRNG that exploits the RowHammer vulnerability in DRAM.
RowHammer is traditionally considered a security threat, where repeated activation of an aggressor row causes unintended bit-flips in adjacent victim rows due to charge leakage and cell-to-cell interference \cite{kim2014flipping,zhou2022lt}. However, we repurpose this phenomenon for privacy. Due to variations in temperature, process nodes, and electronic noise, the exact location and frequency of RowHammer-induced bit-flips are inherently unpredictable, especially under controlled hammering conditions. Once the sense amplifier samples cells near $\frac{V_{DD}}{2}$, the outcomes are affected by thermal noise and become truly random.

EIM-TRNG applies a fixed number of RowHammer cycles, i.e., Hammer Counts (HC), to targeted memory rows. After hammering, the affected page contains a mix of fixed and random bit-flips. By extracting these unpredictable flips as a randomness source, we effectively generate a one-time-use encryption key. The original page and its encrypted counterpart can then be shuffled, and the extracted flip locations stored securely as a decryption key. Our framework provides three major advantages: (1) high security through physically unclonable randomness, (2) no hardware modification since the mechanism exploits existing DRAM behavior, and (3) ease of implementation using common DRAM testing frameworks. This approach provides a one-time, high-entropy solution for model protection, offering a compelling new application of RowHammer: transforming a vulnerability into a tool for enhancing DNN privacy. 

The key contributions of this paper are as follows. (1) We propose a new TRNG  mechanism which provides high-quality randomness derived from physical DRAM behavior, ensuring cryptographic-grade security through inherently unpredictable RowHammer-induced bit-flips.
(2) We propose EIM-TRNG which requires no additional hardware modification, making it readily deployable in commodity DRAM chips. 
(3) We perform extensive experimental characterization and analysis on real DDR4 DRAM modules, rigorously evaluating the security, robustness, and randomness quality of the proposed {EIM-TRNG} approach under realistic operational conditions.

\section{Background \& Motivation}\vspace{-0.3em}
\noindent\textbf{DRAM Architecture.} 
A DRAM chip consists of a two-dimensional array of memory cells, structured into smaller segments known as mats within each bank. Contemporary DRAM modules incorporate billions of such cells \cite{hassan2021uncovering,zhou2024assessing}. Each individual bit-cell, as shown in Fig. \ref{DRAM}, comprises a capacitor and an access transistor, where binary data is encoded based on the capacitor's charge state: a fully charged capacitor represents a logic ``1'' while a discharged capacitor corresponds to a logic ``0'' \cite{zhou2022red,zhou2022flexidram}.

\begin{figure}[h]
\begin{center}
\begin{tabular}{c}
\includegraphics [width=0.96\linewidth]{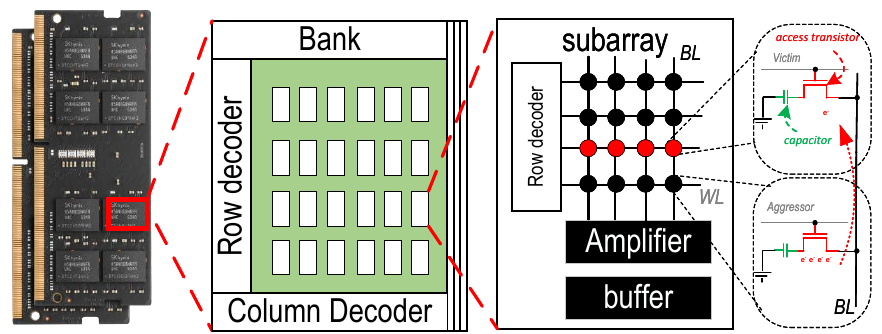}\vspace{-0.4em}
 \end{tabular} \vspace{-0.5em}
\caption{Organization of a DRAM chip.}\vspace{-1em}
\label{DRAM}
\end{center}
\end{figure}

\noindent\textbf{DRAM Timing Constraints.}
The fundamental timing parameter in DRAM is the clock cycle ($t_{CK}$). The Row Access Strobe ({$t_{RAS}$}) defines the period between the activation (\texttt{ACT}) command and the subsequent precharge (\texttt{PRE}) command. Within this duration, the open row undergoes a charge restoration process to maintain data integrity. The Row Precharge Time (\textbf{$t_{RP}$}) determines the minimum time required between a 	\texttt{PRE} and a new \texttt{ACT} command. This interval is crucial for closing an active WL and resetting the bitlines (BLs) to the precharge voltage of $\frac{V_{DD}}{2}$. Additionally, DRAM retention time denotes the maximum period a cell can reliably store data before requiring a refresh operation. The Refresh Window ($t_{REFW}$) specifies the time frame within which all cells must be refreshed to prevent data corruption.

\noindent\textbf{RowHammer Attack.}
Kim et al. \cite{kim2014flipping} conducted a seminal study on RowHammer–induced bit‑flips in DDR3 memory, demonstrating that nearly 85\% of tested modules were vulnerable. Consequently, early RowHammer research focused predominantly on DDR3 systems \cite{seaborn2015exploiting}. With the introduction of DDR4, researchers anticipated improved resilience and several studies have since assessed its robustness \cite{tuugrul2025understanding,olgun2023dram,zhou2024dram}. More recently, DDR5 modules have shown even greater resistance to RowHammer attacks; however, Gloor et al. \cite{gloor2025refault} have just demonstrated successful exploits against two DDR5 devices from  major DRAM manufacturers.

\noindent\textbf{RowHammer Mitigation Strategies.}
Hardware-based countermeasures against RowHammer attacks can be broadly categorized into two approaches: (1) \textit{victim-centric} methods that utilize probabilistic refreshing (e.g., PRA \cite{kim2014architectural}, PARA \cite{kim2014flipping}), and (2) \textit{aggressor-centric} techniques that track activation counts (e.g., U-TRR \cite{hassan2021uncovering}, Hydra \cite{qureshi2022hydra}, TWiCe \cite{lee2019twice}, Graphene \cite{park2020graphene}).
System manufacturers predominantly adopt mechanisms that actively detect RowHammer conditions and mitigate them through increased refresh rates or activation-based tracking. However, these approaches necessitate additional hardware to monitor row activations \cite{frigo2020trrespass,kim2014architectural,qureshi2022hydra,zhou2023dnn,ranyang2023ppim} and maintain activation records using fast-access memory, such as SRAM \cite{lee2019twice} or CAM \cite{park2020graphene}. When a row's activation count reaches the predefined maximum activation count (MAC), the memory controller refreshes the vulnerable row \cite{frigo2020trrespass}.
The JEDEC standard defines three possible configurations for the MAC threshold: (1) \textit{unlimited}, if the DRAM module is classified as RowHammer-resistant; (2) \textit{untested}, if no post-manufacturing validation has been conducted; and (3) $T_{MAC}$, indicating the specific activation threshold before a refresh is required (e.g., 1M activations). A study \cite{frigo2020trrespass} has disclosed that, regardless of the manufacturer, most DDR4 modules declare an \textit{unlimited} MAC value \cite{canpolat2025chronus,zhou2024novel}.

\noindent\textbf{DRAM-based TRNGs.} TRNGs are crucial for cryptographic security, scientific simulations, and industrial testing. Unlike Pseudo-Random Number Generators (PRNGs), TRNGs produce unpredictable values by utilizing physical entropy sources such as electronic noise and quantum effects. Previous works, including DTRNG \cite{humood2021dtrng}, DRNG \cite{eckert2017drng}, FracDRAM \cite{gao2022fracdram}, and QUAC-TRNG \cite{olgun2021quac}, have explored DRAM as a viable entropy source, providing a high-throughput, low-cost, and easily deployable alternative to dedicated TRNG hardware. For instance, DRNG~\cite{eckert2017drng} captures randomness by utilizing DRAM's unpredictable startup behavior. At system startup, the initial state of DRAM cells is influenced by uncontrolled factors such as floating BLs and minor manufacturing discrepancies, resulting in genuinely random voltage patterns. Building upon this, DTRNG\cite{humood2021dtrng} generates randomness through a controlled ``weak write'' process, wherein applying reduced WL voltage causes partial and inconsistent charge transfers. These incomplete transfers introduce significant variability in the cell's final voltage, offering an easily accessible source of entropy without necessitating hardware modifications. FracDRAM~\cite{gao2022fracdram} expands upon this methodology through its specialized ``Frac'' operation, composed of sequential \texttt{ACT} and \texttt{PRE} commands. This operation deliberately places cells into a finely tuned, metastable state near $\frac{V_{DD}}{2}$, exploiting subtle differences in transistor characteristics and capacitances resulting from process variations. Finally, QUAC-TRNG~\cite{olgun2021quac} adopts a similar approach by simultaneously activating and interrupting multiple rows in a QUAC operation, inducing controlled charge sharing among DRAM cells. This results in unpredictable intermediate voltages that, when sampled by sense amplifiers, yield robust, high-quality random bit sequences. QUAC-TRNG \cite{olgun2021quac} operates similarly, utilizing the instability in DRAM cells. Multiple rows in a DRAM segment are activated and interrupted, causing charge sharing and resulting in a bitline voltage near $\frac{V_{DD}}{2}$. The sense amplifier struggles to amplify this small voltage difference, yielding a random value due to thermal noise. QUAC-TRNG performs three steps: initializing the DRAM segment, executing the QUAC operation, and using the SHA-256 hash function to generate high-quality 256-bit random numbers, effectively extracting entropy from DRAM.

\section{Proposed Framework}
\subsection{Threat Model}
In this work, we consider a threat model where the attacker’s primary goal is to extract or reverse-engineer the weights of a deployed DNN model. The attackers often perform side-channel attacks to extract the DNN architectural information~\cite{hu2020deepsniffer, batina2019csi}. Once the architecture is leaked, the attackers can adopt additional side-channel analysis to extract additional model weight parameter information~\cite{rakin2022deepsteal,ahmed2024deep}. Our goal in this work is to obfuscate the model weight parameters; that is, if the attacker performs advanced side-channel attacks to leak DNN model architecture and weights, the recovered version of the model would be completely obfuscated. 

From the hardware perspective, we assume that the attacker has partial low-level system access. Specifically, they have the ability to monitor and infer the physical memory locations that store the DNN weights. For example, they can leverage techniques such as exploiting large pages (e.g., Transparent Huge Pages~\cite{saxena2023pt}, etc.) or performing hardware-based side-channel attacks to map virtual addresses to physical memory addresses. With these capabilities, the attacker can observe which physical pages are used to store weight parameters, even without directly modifying the contents. Moreover, we assume that the adversary can inspect the encrypted or encoded contents of the memory pages allocated to weights. They do not need full root-level access but may gain insights through compromised drivers, kernel modules, or malicious cloud co-tenants in a multi-tenant environment~\cite{rakin2022deepsteal,hu2020deepsniffer}. 

Our objective as a defender is to obfuscate the DNN weight distribution in the memory pages to make them practically useless for an attacker, thus protecting the IP of the model.
The combined threat model reflects a powerful adversary capable of observing DNN weight storage both from software and hardware vectors. 
Our proposed EIM-TRNG framework is designed to defend against such attackers by embedding unpredictable, one-time-use randomness at the hardware level, preventing successful weight extraction or reconstruction.

\begin{figure}[t]
\begin{center}
\begin{tabular}{c}
\includegraphics [width=0.99\linewidth]{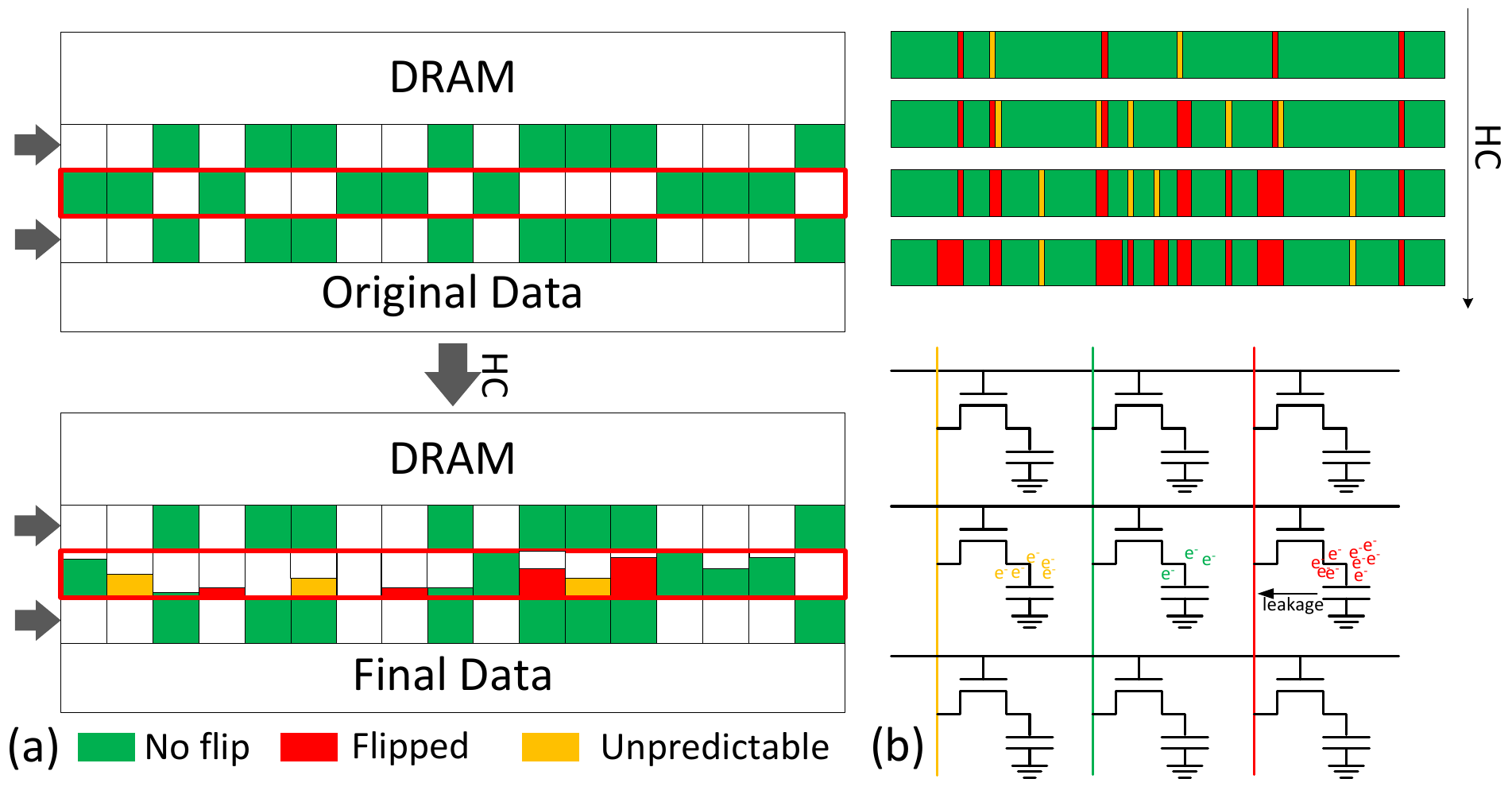}\vspace{-0.4em}
 \end{tabular} \vspace{-0.5em}
\caption{(a) Prescribing RowHammer by increasing HC to create a metastability state in DRAM cells, (b) Leaking the charge via RowHammer to generate a TRN.}\vspace{-2em}
\label{RH}
\end{center}
\end{figure}
\subsection{TRNG Procedure}
In this framework, we use RowHammer to generate TRNs by inducing bit flips in DRAM cells. Fig. \ref{RH} illustrates the fundamental relationship between the RowHammer and the TRNG using DRAM cells. As discussed, RowHammer is a reliability vulnerability in DRAM devices, where repeated activation or hammering of a particular memory row induces unintended electrical disturbances in neighboring rows, causing charge leakage in their memory cells. This leakage alters the stored charges within the affected cells, resulting in data corruption commonly referred to as bit-flips. Specifically, cells that initially store charge levels that represent either logical `1' (full charge close to $\frac{V_{DD}}{2}$) or logical `0' (no charge) can leak toward an intermediate charge level around $\frac{V_{DD}}{2}$. In this state, cells enter a metastable condition, in which the DRAM sense amplifier struggles to resolve the stored value reliably. Due to thermal noise and inherent electrical fluctuations within the memory array, the bit state of these metastable cells becomes inherently unpredictable, providing a natural source of physical randomness.

Importantly, not all DRAM cells exhibit the same susceptibility or behavior under Rowhammer attacks. As shown in Fig. \ref{RH}(b), due to manufacturing process variations (commonly referred to as ``process variability"), individual cells differ significantly in their electrical characteristics, such as capacitance, transistor threshold voltage, and charge retention properties. Consequently, the number of activations (HCs), required to drive a cell's charge toward the unstable $\frac{V_{DD}}{2}$ state varies widely from cell to cell. Some cells, more susceptible due to their physical attributes, require fewer HC to flip their bit states consistently. These cells are identified as ``fixed" or predictable flipped cells. However, other cells demonstrate unpredictable behavior across repeated RowHammer experiments, as they may flip inconsistently or randomly at similar HC levels, making their outcomes practically impossible to forecast.

Fig. \ref{RH}(a) visually distinguishes these categories clearly. For different HCs, the DRAM page data transitions from an initial ``Original Data" state to a corrupted ``Final Data" state containing two distinct categories of bit-flips: fixed flips and unpredictable flips. Regardless of the HC value applied, the affected DRAM region consistently includes a subset of cells that reliably flip (fixed flips) and another subset exhibiting highly inconsistent flipping patterns (unpredictable flips). This second subset, the unpredictable cells, is crucial for randomness generation, as their inconsistent flipping behavior constitutes a robust physical entropy source. Such intrinsic entropy is essential for cryptographic applications, security protocols, and other scenarios requiring true randomness rather than pseudo-randomness derived algorithmically.

Practically, the framework leverages these unpredictable cells by conducting multiple RowHammer experiments under carefully controlled conditions and identifying bits whose flipping behavior cannot be consistently reproduced. Once these unpredictable bits are identified, the framework extracts them as a random bit sequence. Typically, multiple bit segments (e.g., 256 bits) from these unpredictable cells are combined to form cryptographic keys or random numbers stored securely, such as in SRAM, for rapid subsequent access. This approach exploits the natural randomness stemming directly from device-level phenomena rather than relying on computational methods, thus ensuring true unpredictability.
In conclusion, the relationship between RowHammer and TRNG lies fundamentally in exploiting DRAM's inherent physical variability and metastable behavior under electrical disturbances. By deliberately causing charge leakage toward an intermediate unstable state and recognizing that individual cell responses vary significantly due to process variations and thermal noise, this method transforms a reliability threat into a valuable entropy source. The resulting unpredictability, extracted from these metastable cells, fulfills the stringent criteria for generating TRNs, greatly enhancing the security and authenticity of cryptographic applications and secure hardware designs.

\subsection{EIM-TRNG}

\begin{figure}[t]
\centering
\includegraphics[width=1.05\linewidth]{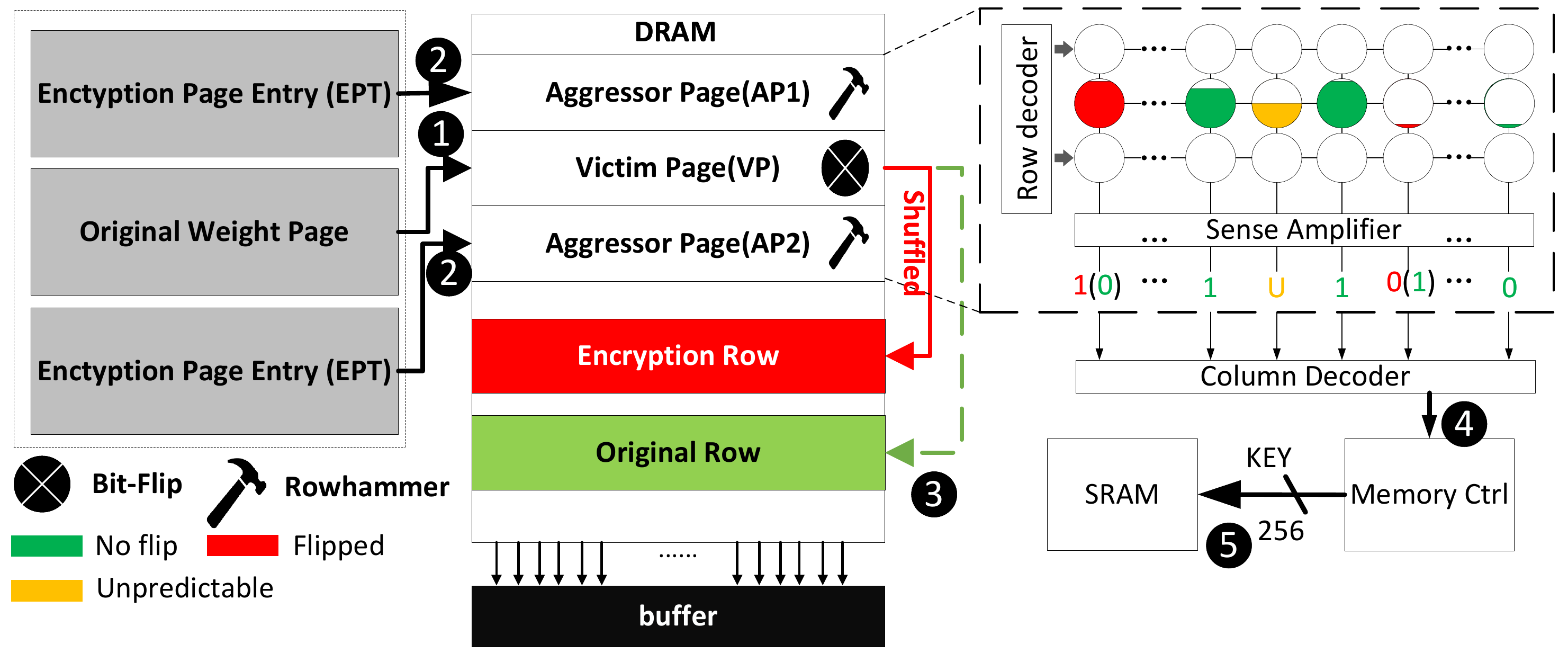}
\caption{Overview of the EIM-TRNG and encrypted weight storage framework. (1) Write Original Weight Page as VP. (2) Write EPT to neighboring rows of VP as AP1 \& AP2. (3) Write weight data to Encryption Row instead of Original Row according to the Flipped VP. (4) Extract the sets including unpredictable bits. (5) Store the 256-bit data as password to SRAM.} \vspace{-1em}
\label{trng-framework}
\end{figure}

\noindent
Fig.~\ref{trng-framework} presents our proposed DRAM-based TRNG framework, tailored for securely concealing DNN model weights. This novel approach leverages the RowHammer effect to produce true randomness derived from physical phenomena. In step \encircle{1}, the \textbf{Original Weight Page} containing the unencrypted DNN model weights is initially loaded into the DRAM row designated as the \textbf{Victim Page (VP)}. This loading step occurs during regular program execution and requires no manual intervention. In step \encircle{2}, adjacent to Victim row, two selected \textbf{Aggressor Pages (AP1 and AP2)} are identified and written. The memory controller can repeatedly activate AP1 \& AP2, deliberately causing electrical disturbances that induce charge leakage in cells of the intermediate {victim page}. As charges leak, certain victim cells go into a metastable state. Due to inherent electrical noise, thermal fluctuations, and variations in DRAM manufacturing processes, the cells entering this metastable state exhibit unpredictability in their resolved bit values, thereby becoming the fundamental source of entropy.

In step \encircle{3}, after a predetermined number of hammering cycles,VP now contains a mixture of unchanged bits, deterministically flipped bits, and unpredictable bit-flips. The flipped VP evolves into a new state termed the \textbf{Encryption Row}. Deterministic flips consistently occur at the same cells due to stable physical properties of these particular DRAM cells, offering limited entropy. Conversely, unpredictable bit-flips exhibit inconsistent flipping patterns across identical RowHammer experiments, driven by subtle variations in cell capacitance, transistor threshold voltages, and environmental conditions. Then the original weight is redirected to the encryption row with the flipped victim page. To effectively harness this inherent unpredictability, our framework executes repeated hammering cycles, collecting data from each iteration to precisely identify the cells whose flipping behaviors are reliably non-deterministic. These unpredictable bits are aggregated to form a robust, physically-derived 256-bit secret key, representing genuine hardware-level entropy suitable for cryptographic applications.

Finally, the memory controller performs a selective read operation utilizing the Column Decoder (step \encircle{4}). This decoder precisely targets and retrieves a contiguous 256-bit segment within the encryption row, optimally chosen based on the density and distribution of unpredictable bit-flips. This selection maximizes the entropy embedded in the cryptographic key. Once the 256-bit segment is selected, it is immediately transferred from DRAM to an on-chip high-speed SRAM buffer (step \encircle{5}). SRAM serves as a low-latency storage medium, providing fast access to the extracted entropy-based key and the encrypted model weights for subsequent cryptographic processing, validation, or authentication. 
Additionally, we propose a data-shuffling mechanism between the original and encrypted pages, further obfuscating positional relationships between the data sets. This shuffling greatly complicates reverse engineering attempts, ensuring that even if attackers access encrypted data, they cannot reliably trace it back to the original weights without possessing the corresponding entropy-based key.

Collectively, this framework effectively leverages DRAM's dual role as both a data storage medium and a source of intrinsic hardware-based randomness. By exploiting RowHammer-induced entropy at the physical level, our approach eliminates reliance on external cryptographic modules or specialized TRNG hardware, thus achieving significant cost and complexity reduction. 
This practical and efficient solution thus provides robust protection of DNN models against unauthorized access, extraction, and tampering at the hardware level.

\section{Experimental Results}
\subsection{Hardware Setup}
To facilitate flexible experimentation on DDR4 memory modules, we construct a customized DRAM attack evaluation framework by significantly adapting the DRAM-Bender platform \cite{olgun2023dram}. Our implementation integrates an in-DRAM compiler API on the host system, allowing the generation and deployment of memory access sequences with fine-grained control. At the hardware level, we utilize the Alveo U200 FPGA board \cite{Alevo}, which interfaces directly with DDR4 DIMMs and serves as the execution engine for test scenarios. As shown in Fig.~\ref{DRAM}, the host machine compiles and transmits DDR4 command traces—corresponding to Algorithms~\ref{alg-1} and~\ref{alg-2}—through the PCIe bus to the FPGA, where they are buffered and executed. This design enables rapid prototyping of complex DRAM access patterns without manual low-level coding. Importantly, our framework accounts for thermal variables via a temperature controller, which plays a critical role in DRAM behavior and is factored into each test configuration to ensure accurate characterization under realistic conditions.

\begin{figure}[h]
\begin{center}
\begin{tabular}{c}
\includegraphics [width=0.72\linewidth]{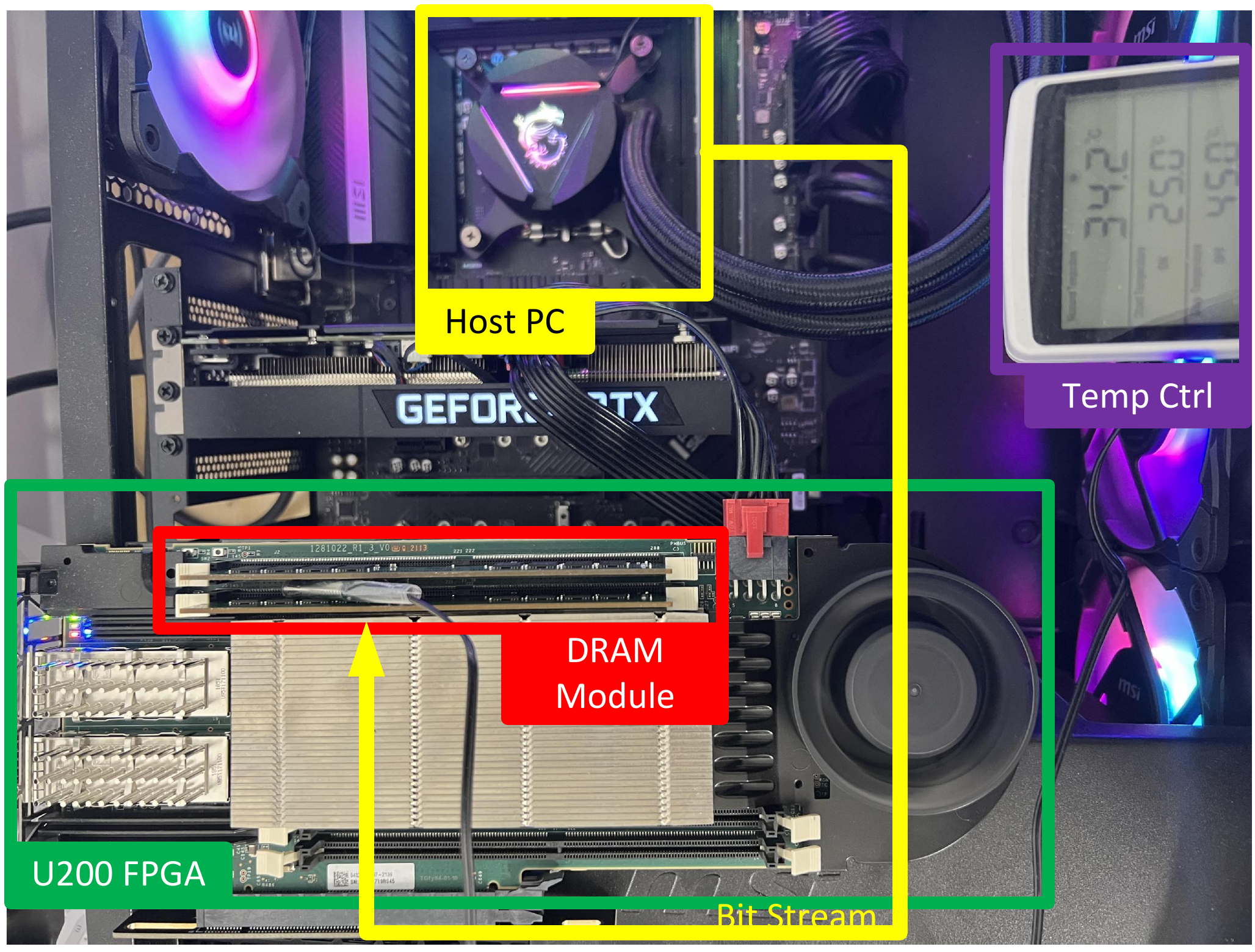}\vspace{-0.4em}
 \end{tabular} \vspace{-0.5em}
\caption{Xilinx Alveo U200 FPGA \& Host PC.}\vspace{-1em}
\label{DRAM}
\end{center}
\end{figure}

\noindent\textbf{Minimizing Interference.} To ensure that we directly observe RowHammer circuit-level bit-flips, DRAM refresh \cite{JEDEC} and rank-level ECC are disabled. However, proprietary RowHammer protection techniques (e.g., Target Row Refresh \cite{frigo2020trrespass,hassan2021uncovering}) exist. 

\noindent\textbf{Tested Commodity DDR4 DRAM Chip.}
We select representative Samsung-manufactured DRAM chips with 16GB density (Frequency: 2400MHz, Die revision: B, Org.: x8, Date: 2053) to profile their cell vulnerabilities.

\subsection{Distribution of Unpredictable Bits}
To evaluate the susceptibility of DRAM cells under RowHammer stress, we first focus on characterizing the number and distribution of affected cells when subjected to a fixed number of activations, referred to as the HCs. A key challenge is that DRAM sense amplifiers produce binary outputs (``0'' or ``1''), and cannot directly sense intermediate voltages such as ($\frac{V_{DD}}{2}$). However, when a DRAM cell's charge level is close to this midpoint, it enters a metastable state. Due to thermal noise and process variations, the sense amplifier may resolve this cell differently in repeated readings, resulting in probabilistic bit-flips. To capture this behavior, we perform repeated experiments with the same HC setting.

\begin{algorithm}[b]
    \caption{\small Bit Flip Probability Analysis via Row Hammering}
    \scalebox{0.65}{
    \begin{minipage}{3\linewidth}
        \begin{algorithmic}[1]
            \State $\textbf{Procedure: \textit{Bit Flip Probability Analysis via Row Hammering}}$
            \State Select victim and aggressor row numbers; specify HC.
            \State Initialize: no\_of\_sets $\rightarrow 10$, iterations per set $\rightarrow 1000$, bitflip\_counts[address] $\rightarrow 0$
            \State $\textbf{For}\hspace{4pt}(s \leq 10)\hspace{4pt}\textbf{do}$
                \State \hspace{16pt}$\textbf{For}\hspace{4pt}(i \leq 1000)\hspace{4pt}\textbf{do}$
                    \State \hspace{32pt}Write data to aggressor and victim rows
                    \State \hspace{32pt}$\textbf{Repeat}\hspace{4pt}hammer\_count\hspace{4pt}\textbf{times}$
                        \State \hspace{48pt}Hammer Aggressor Row 1
                        \State \hspace{48pt}Precharge bank
                        \State \hspace{48pt}Hammer Aggressor Row 2
                        \State \hspace{48pt}Precharge bank
                    \State \hspace{32pt}$\textbf{End Repeat}$
                    \State \hspace{32pt}Read victim row and compare with original data
                    \State \hspace{32pt}$\textbf{If}\hspace{4pt}bit\hspace{4pt}flip\hspace{4pt}occurs$
                        \State \hspace{48pt}Increment flip\_count[address]
                \State \hspace{16pt}$\textbf{End For}$
                \State \hspace{16pt}Store addresses with $0 < flip\_count < 1000$ in vulnerable\_set
            \State $\textbf{End For}$
            \State Identify intersection of all vulnerable\_set as consistent\_vulnerable\_cells
            \State $\textbf{For each cell} \in$ consistent\_vulnerable\_cells
                \State \hspace{16pt}Compute average flip count over 10 sets
                \State \hspace{16pt}Calculate flip probability $= \frac{\text{average count}}{1000}$
            \State Output: list of consistently vulnerable bit cells and their probabilities
            \State \textbf{end} $\textbf{Procedure}$
        \end{algorithmic} 
    \end{minipage}
    }
    \label{alg-1}
\end{algorithm}

Algorithm~\ref{alg-1} outlines the procedure for analyzing bit-flip probabilities induced by RowHammer operations. Initially, the victim and aggressor rows, along with a specified HC, are selected. The analysis is conducted over multiple sets (ten sets by default), each containing 1000 repeated hammering iterations. For each iteration, data is written to the chosen aggressor and victim rows, after which the aggressor rows are repeatedly activated (hammered), followed by bank precharges, to induce charge leakage in victim cells. After the hammering sequence, the victim row is read and compared against the original data to detect bit-flips. Any detected bit-flip increments a counter associated with the affected cell's address. At the end of each set, addresses that experienced inconsistent flipping (the number of flipping between 0 and 1000) are recorded in a vulnerable cell set. The intersection of these vulnerable sets across all ten experiments identifies consistently vulnerable cells. For these identified cells, the average flip count is calculated, allowing computation of their flip probability by dividing this average by the total iterations (1000). The algorithm ultimately outputs a list of consistently vulnerable DRAM cells along with their flip probabilities, enabling detailed characterization of RowHammer-induced vulnerabilities.

\begin{figure}[t] 
\centering
\includegraphics [width=1\linewidth]{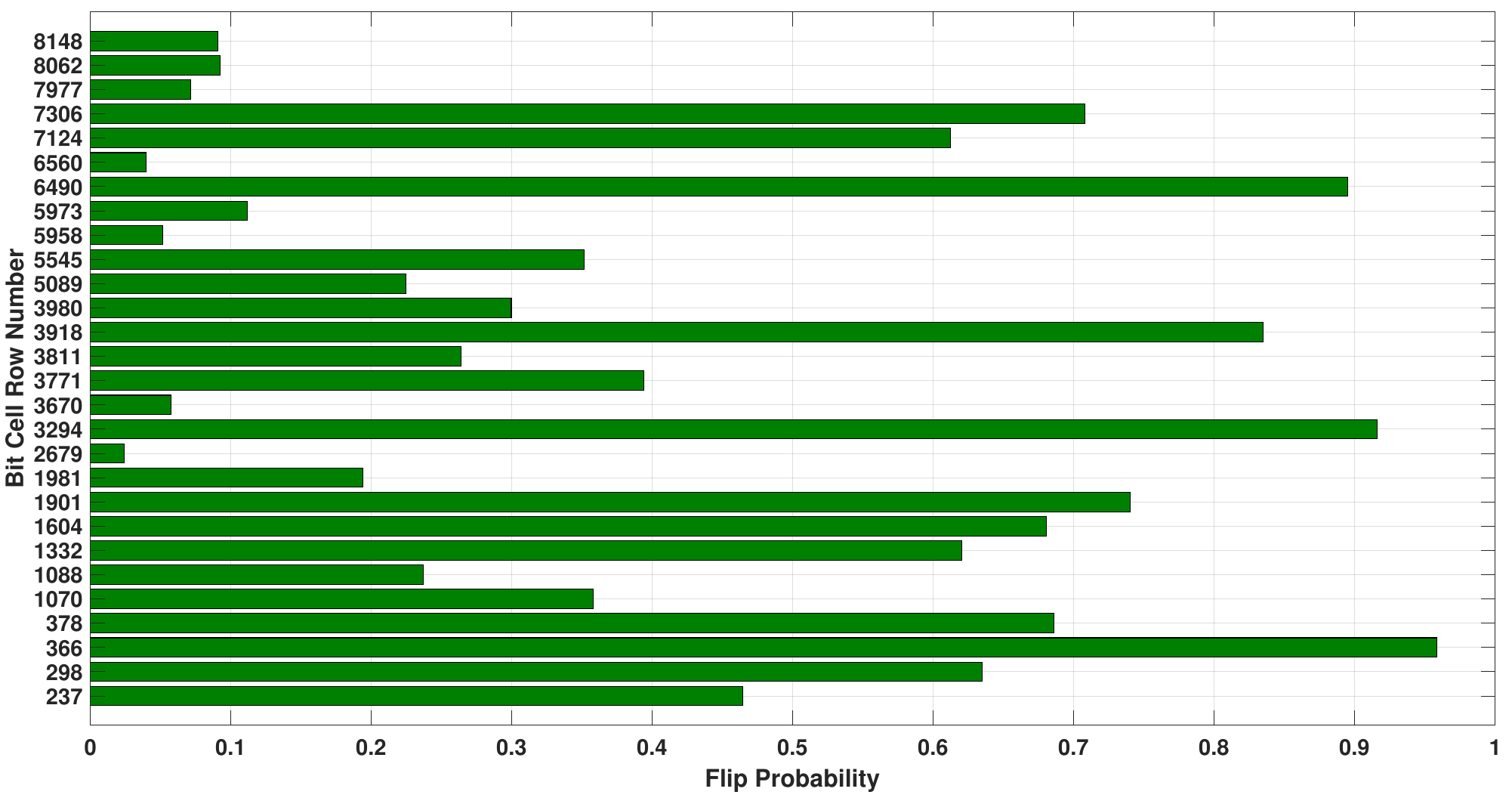}
\vspace{-2em}
\caption{Bit-flip probabilities of selected DRAM cells after applying RowHammer with a fixed hammer count.}
\vspace{-1em}
\label{Probability_distribution}
\end{figure}

Fig. \ref{Probability_distribution} illustrates the bit-flip probabilities for a selected set of DRAM cells subjected to a fixed HC (1M). Each DRAM cell, identified on the vertical axis by its specific row number, exhibits a unique probability of bit-flipping, ranging horizontally from 0 (no flips observed) to 1 (all flips observed). The distribution clearly reveals significant variation among these cells, underscoring the inherent unpredictability arising from physical differences such as transistor threshold variability, capacitance variations, and manufacturing inconsistencies across cells within the same memory array.  Specifically, several cells demonstrate consistently high flip probabilities, approaching 1, indicating a near-certain likelihood of flipping under repeated RowHammer stress. For instance, rows such as 366, 3294, and 6490 exhibit flip probabilities approaching unity, indicating highly vulnerable cells with consistently unstable behavior. In contrast, other cells like rows 3670 and 2679 demonstrate significantly lower flip probabilities—below 0.1—suggesting greater stability and resistance to charge leakage. Between these cells, a large number of cells display intermediate flip probabilities ranging from 0.2 to 0.8. These cells represent a partially predictable yet inherently unstable category, whose charge retention characteristics fluctuate under repeated RowHammer stress, contributing valuable entropy for TRNG.

The presence of cells spanning the entire probability spectrum highlights the complexity and richness of DRAM-based randomness sources. Cells with intermediate and low probabilities are particularly valuable in security applications, as their unpredictable and probabilistic flipping behavior can be directly utilized to generate TRNs or cryptographic keys. Importantly, the observed variation in flip probability among different cells ensures robust entropy generation, as the diverse cell behaviors provide a substantial source of randomness, making attempts at predicting or replicating these patterns exceedingly difficult. 
By selectively combining cells of varying vulnerability and unpredictability, this method enables the construction of strong cryptographic keys, thereby enhancing data protection and reinforcing hardware security measures against potential attacks.

\subsection{Performance of Charge Leakage in Different Cells} 
Algorithm~\ref{alg-2} describes a systematic approach for analyzing DRAM cell vulnerability by examining the relationship between bit-flip occurrences and varying RowHammer activation counts. Initially, the victim and aggressor rows are selected, alongside a range of HCs defined by starting and ending values with increments of 1000 activations per step. For each specified HC, the algorithm performs 1000 iterations, wherein the aggressor rows are repeatedly activated with intervening precharges, deliberately inducing electrical disturbances into the victim row cells. After each iteration, data from the victim row is read and compared against the original values to detect bit-flips, incrementing corresponding counters upon detection. Once the iterations conclude for each HC, the algorithm records cells exhibiting partial flipping behavior—cells that flip occasionally but not consistently across all trials. The intersection of these vulnerable cell sets across different HCs identifies consistently unstable cells. Finally, the algorithm outputs these consistently vulnerable cells along with detailed bit-flip statistics corresponding to each HC, thereby allowing comprehensive characterization of DRAM susceptibility and offering valuable insight into the reliability and randomness properties of DRAM cells under varying RowHammer stress conditions.

\begin{algorithm}[h]
    \caption{\small Bit Flip Count Analysis Based on Varying Row HC}
    \scalebox{0.65}{
    \begin{minipage}{3\linewidth}
        \begin{algorithmic}[1]
            \State $\textbf{Procedure: \textit{Bit Flip Count Analysis Based on Varying Row HC}}$
            \State Select victim and aggressor rows; 
            \State set $start\_hammer\_count$, $end\_hammer\_count$, and $step = 1000$
            \State Initialize: $iterations\_per\_hammer \rightarrow 1000$, $bitflip\_count[address] \rightarrow 0$
            \For{$hammer\_iter = start\_hammer\_count$ \textbf{to} $end\_hammer\_count$ \textbf{step} $1000$}
                \For{$i = 1$ \textbf{to} $iterations\_per\_hammer$}
                    \State Write data to aggressor and victim rows
                    \For{$j = 1$ \textbf{to} $hammer\_iter$}
                        \State Hammer Aggressor Row 1
                        \State Precharge bank
                        \State Hammer Aggressor Row 2
                        \State Precharge bank
                    \EndFor
                    \State Read victim row and compare with original data
                    \If{bit flip occurs}
                        \State Increment $bitflip\_count[address]$
                    \EndIf
                \EndFor
                \State Store addresses with $0 < bitflip\_count < 1000$ in $vulnerable\_set[hammer\_iter]$
            \EndFor
            \State Identify intersection of all $vulnerable\_set[hammer\_iter]$ as $consistent\_vulnerable\_cells$
            \State Output: list of consistently vulnerable bit cells and their bit flip count per each HC
            \State \textbf{end} $\textbf{Procedure}$
        \end{algorithmic}
    \end{minipage}
    }
    \label{alg-2}
\end{algorithm}

\begin{figure*}[t] 
\centering
\includegraphics [width=0.93\linewidth]{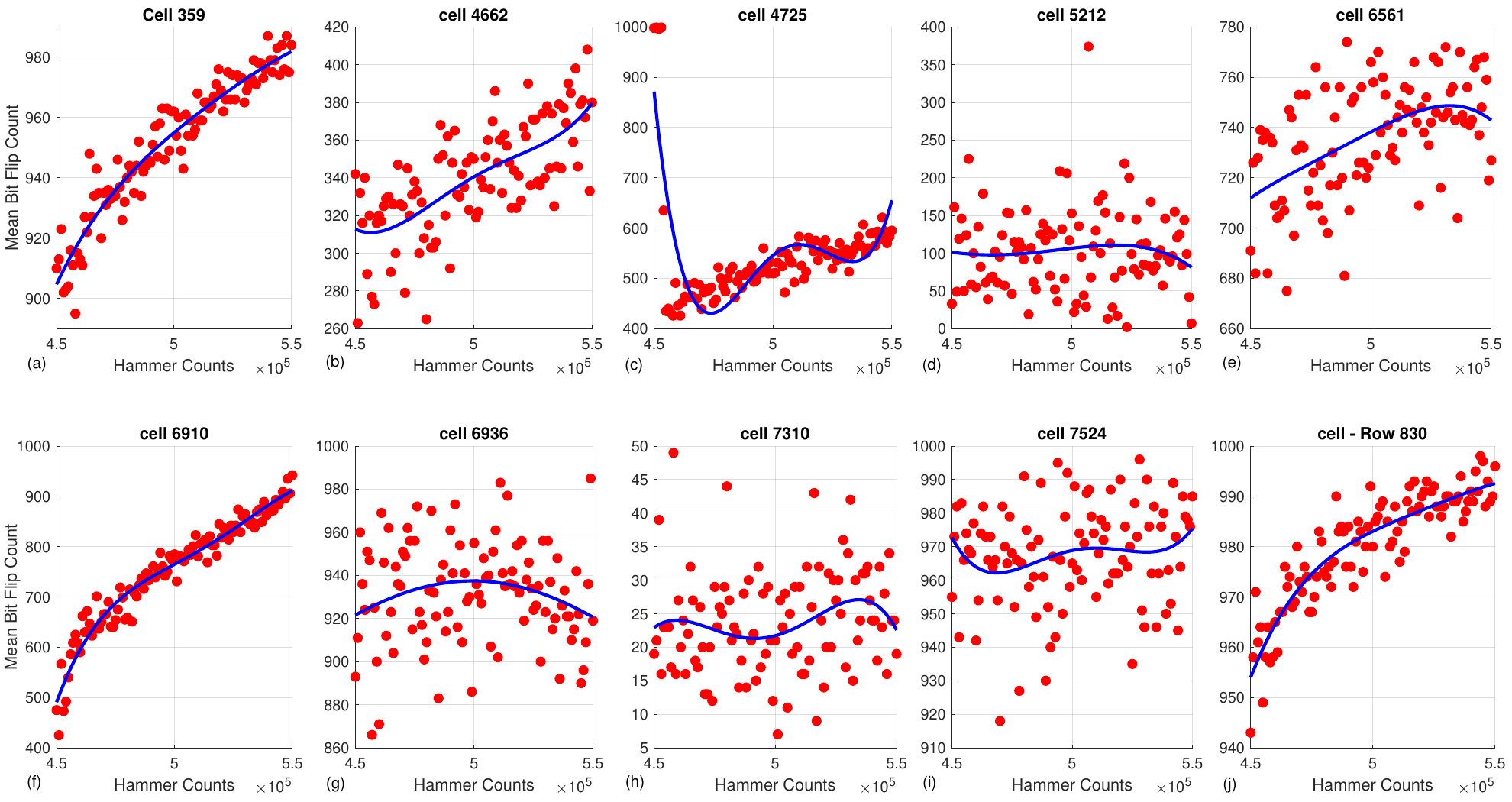}
\vspace{-1em}
\caption{Bit-Flip behavior of selected DRAM cells under RowHammer Attacks with the mean bit-flip count for ten individual DRAM cells across varying HCs. Each subplot corresponds to a unique DRAM cell, identified by its row address: {(a)} Cell 359, {(b)} Cell 4662, {(c)} Cell 4725, {(d)} Cell 5212, {(e)} Cell 6561, {(f)} Cell 6910, {(g)} Cell 6936, {(h)} Cell 7310, {(i)} Cell 7524, and {(j)} Cell from Row 830.}
\vspace{-1em}
\label{Probability_cells}
\end{figure*}

The plots presented in Fig.~\ref{Probability_cells} illustrate the relationship between the HC and the mean bit-flip occurrences across multiple DRAM cells. Each plot shows a distinct DRAM cell's sensitivity to RowHammer effects, capturing the variability in cell responses due to inherent manufacturing differences and physical variability. As observed, the behavior of these cells varies significantly. For example, cells such as 359, 6936, 7524, and the cell from Row 830 demonstrate very high sensitivity to RowHammer, with their mean bit-flip counts rapidly approaching the maximum count of 1000 flips as the HC increases. This indicates consistent and predictable flipping behavior, rendering these cells reliably vulnerable under repeated RowHammer stress. Conversely, cells like 4662 and 5212 display moderate bit-flip susceptibility, showing a gradual increase in flip count with increasing HCs but remaining far below the maximum. This moderate flipping frequency highlights the stochastic and probabilistic nature of their behavior, likely due to charge instability near $\frac{V_{DD}}{2}$ levels caused by intermittent charge leakage. Such cells contribute significantly to randomness, providing substantial entropy for TRNG. Other cells, notably cell 7310, exhibit low and sporadic bit-flip behavior, characterized by a very limited number of flips even at higher HCs. These cells represent a category with extremely low sensitivity, potentially flipping only under extreme conditions, thus offering limited utility as randomness sources. Cell 6561 and cell 4725 exhibit intermediate but non-monotonic behaviors, indicating complex internal electrical characteristics. Such non-linear and unpredictable patterns further demonstrate the intricate interplay between physical cell characteristics and the RowHammer phenomenon.

\begin{table*}[t]
\centering
\caption{Quantitative and qualitative comparison of DRAM-based TRNG frameworks.}
\vspace{-1em}
\label{tab:trng-summary}
\scriptsize
\resizebox{\textwidth}{!}{
\begin{tabular}{|l|l|c|c|c|c|c|c|c|l|}
\hline
\textbf{TRNG} & \textbf{Entropy Mechanism} & \textbf{Latency} & \textbf{Thpt} & \textbf{Power} & \textbf{Mod} & \textbf{WW} & \textbf{Preq} & \textbf{OTU} & \textbf{Security Summary} \\
 &  & ($\mu$s) & (Mb/s) & (pJ/bit) & HW &  &  &  &  \\
\hline
\textbf{EIM-TRNG} & RH-induced metastability & $>$1500 & 0.1--1 & $>1.2\times10^5$ & \xmark & \xmark & \xmark & \cmark & No buffer/key; entropy embedded in data \\
\hline
QUAC-TRNG~\cite{olgun2021quac} & Quad-row ACT + bitline noise & 1.94 & 3440 & $\sim$1.5 & \xmark & \xmark & \cmark & \xmark & Entropy exposed; row pattern inference possible \\
\hline
FracDRAM~\cite{gao2022fracdram} & Partial precharge + sense amp & 0.0175 & 15--20 & 20--40 & \xmark & \xmark & \cmark & \xmark & Flip bias predictable under modeling \\
\hline
DTRNG~\cite{humood2021dtrng} & Weak WL write randomness & 0.00025 & 4000 & 50 & \cmark & \cmark & \cmark & \xmark & Modifiable but vulnerable to profiling \\
\hline
DRNG~\cite{eckert2017drng} & DRAM power-on startup bits & $>$6400 & 0.1--0.5 & Static & \xmark & \cmark & \cmark & \cmark & Repeatable under reboot; needs correction \\
\hline
D-RaNGe~\cite{kim2019d} & Activation failure + retention entropy & 1.5 & 7480 & $\sim$2.0 & \xmark & \xmark & \cmark & \xmark & Requires special DRAM cells; good randomness and high throughput \\
\hline
DR-STRaNGe~\cite{bostanci2022drstrange} & Retention + buffered sampling & 5–10 & 300--500 & $\sim$100 & \xmark & \xmark & \cmark & \xmark & System-friendly integration with memory access fairness \\
\hline
DRAM-Latency~\cite{rahman2019dramlatency} & Reduced tRCD-induced failures & 0.8–2 & 10--50 & $\sim$20 & \xmark & \xmark & \cmark & \xmark & Low-cost entropy source from timing violations \\
\hline
\end{tabular}}
\vspace{-1em}
\footnotesize \textbf{Thpt:} Throughput, \textbf{WW:} Weak Write, \textbf{Mod:} Hardware Modification, \textbf{Preq:} Prerequisite, \textbf{OTU:} One-Time Use.
\vspace{-1em}
\end{table*}

\subsection{Evaluation of Model Leakage Attack in DNN} 
This section evaluates the performance of the framework in protecting a stolen DNN weight parameter obfuscated by EIM-TRNG. To validate that our proposed method can successfully obfuscate a DNN model and make the recovered version practically in-effective, we evaluate the method using two key criteria: $(i)$ For a given DNN model, our protection should demonstrate that after swapping $N$ number of secret rows with original rows the DNN models accuracy reaches close to a random guess level, i.e., the model’s accuracy drops to $\frac{100}{c}\%$, where $c$ is the number of output class. We can successfully generate $N$ numbers of random numbers, and our SRAM design has the capability to store them. $(ii)$ Once an attacker retrieves the obfuscated model, they should not be able to trivially recover the original performance of the model by performing gradient descent with a limited amount of available training data. 

We used two representative models, ResNet-20 and ResNet-56, to perform the above two tests. Our evaluation on the ResNet-20 and ResNet-56 models shows that they require 50 and 80 swaps, respectively, between the pagetable entries to ensure the model performance gets close to 10\%  (random guess level for the CIFAR-10 dataset~\cite{krizhevsky2014cifar}). The random number range (e.g., 50-80) is small and can be easily generated and stored successfully on the SRAM memory. As for the second test, we took both the obfuscated model and trained it with 1\% of training samples, assuming an attacker who attempts to steal the model parameters does not have a large dataset and training facilities available to them. Our evaluation shows that with 1 \% training data, it is practically impossible to recover the obfuscated DNN model's functionality at all. We observe that training with 1 \% data on our obfuscated model resulted in the exact same level of accuracy recovery as taking randomly initialized weights. \emph{This validates that stealing this obfuscated model version is equivalent to an attack, as taking a random model with no utility.}

\subsection{Comparison with Other TRNG Frameworks}

Table~\ref{tab:trng-summary} provides a comprehensive comparison of state-of-the-art DRAM-based TRNG frameworks across multiple dimensions, including entropy mechanism, performance, hardware requirements, and security robustness. While \textbf{EIM-TRNG} incurs significantly higher latency (greater than 1500~$\mu$s) and energy consumption ($1.2 \times 10^5$~pJ/bit) compared to prior work such as FracDRAM~\cite{gao2022fracdram}, QUAC-TRNG~\cite{olgun2021quac}, and DTRNG~\cite{humood2021dtrng}, its energy cost remains notably lower than that of DRNG~\cite{eckert2017drng}, which requires activation of the entire DRAM array to extract startup values. This highlights a distinct trade-off: \textbf{EIM-TRNG} prioritizes enhanced security and ease of deployment on unmodified hardware. Moreover, since our framework supports offline generation, the latency and energy overhead can be amortized or hidden from runtime performance, making it practical for secure AI model protection scenarios.Unlike FracDRAM and QUAC-TRNG, which require custom modifications to DRAM timing behavior or activation sequences, EIM-TRNG operates on commodity DRAM chips without any hardware or protocol changes, making it easily adoptable in real-world systems. 

In contrast to DTRNG and DRNG, which depend on weak write operations or startup state randomness that can be partially characterized or reproduced under specific conditions, EIM-TRNG leverages RowHammer-induced metastability. The resulting randomness arises from deep physical-level factors such as thermal noise, cell topology, and manufacturing variation—making it inherently harder to predict, even under repeated runs.We also compare our method with more recent frameworks. 

D-RaNGe~\cite{kim2019d} achieves high throughput by targeting DRAM regions with activation failures, offering low-latency entropy extraction. However, it depends on profiling special cells and cannot guarantee universal deployability across DRAM modules. DR-STRaNGe~\cite{bostanci2022drstrange} proposes a system-friendly TRNG design that improves integration fairness and hides latency via buffering, though it does not encode entropy into data. DRAM-Latency TRNG~\cite{rahman2019dramlatency} introduces read errors through reduced timing margins, providing fast, low-cost randomness but requiring careful control of DRAM access parameters. Compared to these, EIM-TRNG directly embeds entropy into encrypted memory, providing security benefits even under adversaries with full DRAM visibility. 

Crucially, from a security perspective, EIM-TRNG stands out by embedding entropy directly into the protected data itself. This design eliminates the need for temporary keys or buffer storage, which could otherwise be observed or extracted by an adversary. Even under a strong threat model with full DRAM visibility, the framework ensures only obfuscated data is exposed. Without knowledge of the unpredictably flipped bits or access to the internal mapping, reconstructing the original DNN weights becomes infeasible. Moreover, since the entropy is directly embedded within the encoded page, traditional key-recovery methods are ineffective. EIM-TRNG supports configurable HC values to generate distinct TRN patterns based on the physical locations of unpredictable cells. This cell-specific behavior, driven by manufacturing variability and environmental noise, enhances the uniqueness of each TRN instance. While this approach may result in higher latency and energy cost, it remains a robust and deployable solution for embedding hardware-level randomness into DNN data encoding across a variety of secure AI applications.


\section{Conclusion \& Future Work}
By leveraging RowHammer-induced charge leakage, our approach introduces a charge-leakage randomization mechanism that transforms predictable memory behavior into a source of true randomness. Although our method incurs greater latency and power consumption compared to prior TRNG frameworks such as FracDRAM, QUAC-TRNG, DRNG, and DTRNG, it offers stronger security guarantees and ease of implementation on unmodified commodity DRAM chips. Our EIM-TRNG not only generates unpredictable outputs but also integrates them directly into the model’s weight storage, effectively concealing sensitive data even when the attacker has full access to the DRAM content. In doing so, it obfuscates the underlying model while providing no explicit trace of the random key, making reverse engineering significantly more difficult. Through extensive experiments, we analyzed the behavior of DRAM cells under varying RowHammer counts and observed consistent patterns of unpredictable bit-flips tied to physical variability such as process noise, temperature, and layout topology. Our evaluation confirms that a subset of these cells consistently produces non-deterministic results, offering a reliable entropy source for secure applications.

As part of future work, we plan to further investigate the relationship between elevated operating temperatures and reduced hammering thresholds, with the aim of decreasing the number of required RowHammer activations for entropy generation. Additionally, we intend to explore more efficient charge-leakage mechanisms and memory access patterns that could lower energy consumption while preserving randomness quality. Beyond optimization, we are also interested in extending EIM-TRNG to other security applications, such as hardware-level watermarking of AI models, secure federated learning, and integration into trusted execution environments. Ultimately, our work demonstrates a novel path toward embedding robust security mechanisms directly into commodity memory systems, bridging the gap between physical-layer vulnerabilities and practical hardware-based protection for AI workloads.

\bibliographystyle{IEEEtran}
\bibliography{Reference}

\end{document}